\documentclass[12pt]{article}
\usepackage{graphicx}
\usepackage{amsfonts}
\usepackage{amssymb,amsmath}
\usepackage{color}
\usepackage[colorlinks=true,linkcolor=blue,citecolor=blue]{hyperref}

\begin{document}

\begin{titlepage}

\begin{center}

{\Large \bf 
Optimal constraint on $g_{\rm NL}$ from CMB
}

\vskip .45in

{
Toyokazu Sekiguchi$^a$ and
Naoshi Sugiyama$^{a,b,c}$
}

\vskip .45in

{\em
$^a$Department of Physics and Astrophysics, Nagoya University, Nagoya 464-8602, Japan\\
$^b$Kobayashi-Maskawa Institute, Nagoya University, Nagoya 464-8602, Japan \vspace{0.2cm}\\
$^c$ Kavli Institute for the Physics and Mathematics of the Universe,
University of Tokyo, Kashiwa 277-8568, Japan \vspace{0.2cm}
}

\end{center}

\vskip .4in

\begin{abstract}
An optimal method to constrain the non-linearity parameter $g_{\rm NL}$ of the local-type non-Gaussianity from CMB data is proposed.  
Our optimal estimator for $g_{\rm NL}$ is separable and can be efficiently computed in real space.
Combining the exact filtering of CMB maps with the full covariance matrix, our method allows us 
to extract cosmological information from observed data as much as possible 
and obtain a tighter constraint on $g_{\rm NL}$ than previous studies.
Applying our method to the WMAP 9-year data, we obtain the constraint $g_{\rm NL} = (-3.3 \pm 2.2) \times 10^5$, 
which is a few times tighter than previous ones. We also make a forecast for PLANCK data by using the Fisher matrix analysis.
\end{abstract}

\end{titlepage}

\setcounter{page}{1}

\section{Introduction} 
\label{sec:introduction}

One of the most fundamental questions in cosmology is
what is the origin of the primordial fluctuations which seed
the large scale structure in the observed Universe as well
as the anisotropies in the cosmic microwave
background (CMB). Various cosmological observations
consistently show that primordial fluctuations are adiabatic, nearly 
scale-invariant and Gaussian, which is consistent with
a prediction of simple single-field slow-roll inflation models.
On the other hand, there are a variety of models in which
probability distribution of primordial perturbations can significantly 
derivate from Gaussian ones.

Among an infinite number of possibilities for deviation from
Gaussian distributions, we in this paper focus on the local-type non-Gaussianity
\cite{Komatsu:2001rj}, in which the non-Gaussian curvature perturbation $\Phi(\vec x)$
is given as a function of its Gaussian part $\Phi_{\rm G}(\vec x)$ only at the same point, i.e. 
\begin{equation}
\Phi(\vec x)=\Phi_{\rm G}(\vec x)+
f_{\rm NL}\left[\Phi_{\rm G}(\vec x)^2-\langle \Phi_{\rm G}(\vec x)^2\rangle\right]
+g_{\rm NL}\Phi_{\rm G}(\vec x)^3+\cdots, 
\label{eq:NGpert}
\end{equation}
where $f_{\rm NL}$ and $g_{\rm NL}$ are called non-linearity parameters.
This type of non-Gaussianity is of particular interest.
In standard single-field slow-roll
inflation models, amount of this type of non-Gaussianity is too small
to be observed at least in the near future. On the other hand, 
a range of theoretical models based on the inflationary Universe, 
in which multiple degrees of freedom during inflation
contribute to primordial perturbations, can generate a large
local-type non-Gaussianity \cite{Lyth:2005fi}.
Among them, typical examples are 
the curvaton scenario 
\cite{Enqvist:2001zp,Lyth:2001nq,Moroi:2001ct}
and the modulated reheating scenario \cite{Dvali:2003em,Kofman:2003nx}.
Therefore the local-type non-Gaussianity
is a unique probe for these models which may manifest only 
at the very early Universe and very high energy scales.

So far many attempts have been made to detect $f_{\rm NL}$.
Current constraints come from largely two types of observations.
One is the bispectrum of the temperature anisotropy in CMB.
Current data from the WMAP 9-year observation gives a constraint
$-3<f_{\rm NL}<77$ at 95\% confidence level (C.L.) 
\cite{Bennett:2012fp}.\footnote{
While primordial perturbations are consistent with adiabatic ones
and non-Gaussianity in curvature perturbations is discussed in
the most literature, some theoretical models predict isocurvature perturbations 
which can have a local-type non-Gaussianity, e.g., Refs. \cite{Kawasaki:2008sn,Kawasaki:2008pa}.
Current constraints on isocurvature non-Gaussianities are
presented in Refs. \cite{Hikage:2012be,Hikage:2012tf}.
} Another probe is the scale-dependent bias in correlation functions
of massive objects. At present the best constraint
from the scale-dependent bias gives $-37<f_{\rm NL}<25$ at 95\% C.L. 
\cite{Giannantonio:2013uqa}, where angular correlation functions of galaxies and quasars
are used.

While $f_{\rm NL}$, which parameterizes the leading-order non-Gaussian term in Eq.\eqref{eq:NGpert}, 
have already begun to be constrained by various data, there are higher-order terms, 
which remain to be explored more deeply.
Regarding $g_{\rm NL}$, which
is the coefficient of the next-to-leading order term in Eq. \eqref{eq:NGpert}, 
so far several groups \cite{Vielva:2009jz,Smidt:2010ra,Fergusson:2010gn,Hikage:2012bs}
have presented CMB constraints on it.\footnote{
$g_{\rm NL}$ can be also constrained from the scale-dependent biases. For current constraints, 
we refer to Ref. \cite{Giannantonio:2013uqa}.
} 
The authors of Refs .\cite{Smidt:2010ra,Fergusson:2010gn} especially use optimal estimators of $g_{\rm NL}$.
In Ref. \cite{Smidt:2010ra}, the authors present a constraint $-7.4\times 10^5<g_{\rm NL}<8.2\times 10^5$ 
(95\% C.L.) from the WMAP 5-year data.
Using the same data the authors in Ref. \cite{Fergusson:2010gn} obtained 
The resultant constraint is $g_{\rm NL}=(1.6\pm7.0)\times 10^5$ at 1 $\sigma$.
While their estimators are optimal, in actual implementation
several approximations are adopted, which can weaken the resultant constraints.
In particular, so far no constraints have taken into account the inhomogeneity in noise levels and
sky cuts accurately. In addition, multipoles included in the analysis are to some extent limited, i.e. $l\lesssim 500$. Thus, we suspect
cosmological information contained in data may not be fully extracted.
Meanwhile there are models with large $g_{\rm NL}$ which may
be observationally detected 
(See, e.g., Refs. \cite{Enqvist:2008gk,Enqvist:2009eq}).\footnote{
We refer to Ref. \cite{Suyama:2010uj} 
as a review of theoretical models which predict large local-type non-Gaussianities.
}
Therefore it is important to improve constraints on $g_{\rm NL}$
and enable an optimal estimation of it from CMB observations.

In this paper, we discuss a method to derive an optimal constraint on
$g_{\rm NL}$ from CMB observations. 
In particular, we show that an optimal estimator of $g_{\rm NL}$ can be computed 
efficiently in real space. Moreover, our implementation of estimation 
is optimal with the full covariance matrix being taken into account
and the maximum multipole being large enough.
Organization of this paper is as follows. In the next section, we discuss 
a CMB trispectrum generated from $g_{\rm NL}$. In Section \ref{sec:estimator}, 
we present a representation of the optimal estimator for $g_{\rm NL}$ in real space, 
which we compute in this paper.  In Section \ref{sec:result}, after describing details of 
our analysis, we present constraints on $g_{\rm NL}$ from 
the WMAP 9-year data. We also compare our results 
with a forecast based on the Fisher matrix analysis in Section 
\ref{sec:fisher}. The final section is devoted to conclusion.

Throughout this paper, 
we assume a concordance flat power-law $\Lambda$CDM model, 
and the cosmological parameters are fixed to the mean values from 
the WMAP 7-year data alone \cite{Komatsu:2010fb}, 
\begin{equation}
(\Omega_b,\Omega_c, H_0, \tau, n_s, A_s)=
(0.0448, \,0.220, \,71, \,0.088, \,0.963, \,2.43\times 10^{-9}).
\end{equation}
Here, $\Omega_b$ and $\Omega_c$ are respectively the density parameters for 
baryon and CDM, $H_0$ is the Hubble constant in units of 
km/sec/Mpc, $\tau$ is the optical depth of reionization, and
$n_s$ and $A_s$ are respectively the spectral index and amplitude of power spectrum of 
curvature perturbations at a reference scale $k_*=0.002$Mpc$^{-1}$, 
i.e. $P_\Phi(k)=\frac{2\pi^2}{k^3}\frac{25A_s}{9}(\frac{k}{k_*})^{n_s-1}$.

\section{CMB trispectrum from nonzero $g_{\rm NL}$} 
\label{sec:NG}

First let us consider correlation functions of primordial curvature perturbations
$\Phi$ in the local-type non-Gaussianity.
A non-trivial effect of $g_{\rm NL}$ arises in the connected part 
of the four-point correlation function of $\Phi(\vec x)$, or its Fourier dual, the trispectrum.
If $f_{\rm NL}=0$, the connected trispectrum should be given by
\begin{equation}
\langle \Phi (\vec k_1)\Phi (\vec k_2)
\Phi (\vec k_3)\Phi (\vec k_4)\rangle_{\rm conn}
=6g_{\rm NL}\left[P_{\Phi}(k_1)P_{\Phi}(k_2)P_{\Phi}(k_3)+(\mbox{3 perms})
\right](2\pi)^3\delta^{(3)}(\vec k_{1234}),
\label{eq:primt}
\end{equation}
where $\vec k_{i_1\cdots i_n}\equiv \vec k_{i_1}+\cdots+\vec k_{i_n}$.
For simplicity, we in this paper are to constrain $g_{\rm NL}$, assuming
$f_{\rm NL}$ to be zero.

Neglecting the secondary non-Gaussianities
arising from the second- or higher-order cosmological perturbation theory, 
the harmonic coefficients of the CMB temperature anisotropy from 
primordial perturbations $\Phi$ can be given as
\begin{equation}
a_{lm}=4\pi(-i)^l\int \frac{d^3k}{(2\pi)^3}g_l(k)\Phi(\vec k)Y^*_{lm}(\hat k), 
\label{eq:alm}
\end{equation}
where $g_l (k)$ is the temperature transfer function.
The angular power spectrum $C_l$ of the temperature anisotropy, which is defined by 
$\langle a_{lm}{a_{l'm'}}^*\rangle=C_l\delta_{ll'}\delta_{mm'}$,  can be given as
\begin{equation}
C_l=\frac2\pi
\int dk\,k^2
{g_l(k)}^2
P(k).
\end{equation}

We define a reduced CMB trispectrum $t_{l_1l_2l_3l_4}$,\footnote{
Note that our definition of a reduced trispectrum, $t_{l_1l_2l_3l_4}$, is different from the 
one in Ref. \cite{Kogo:2006kh}, $t^{l_1l_2}_{l_3l_4}(L)$.}
\begin{equation}
\langle a_{l_1m_1}a_{l_2m_2}a_{l_3m_3}a_{l_4m_4}\rangle_{\rm conn}
\equiv t_{l_1l_2l_3l_4}\mathcal G^{l_1l_2l_3l_4}_{m_1m_2m_3m_4}, 
\label{eq:tlm4}
\end{equation}
where $\mathcal G^{l_1l_2l_3l_4}_{m_1m_2m_3m_4}$ is defined by
\begin{equation}
\mathcal G^{l_1l_2l_3l_4}_{m_1m_2m_3m_4}\equiv
\int d\hat n \,Y_{l_1m_1}(\hat n)
Y_{l_2m_2}(\hat n)Y_{l_3m_3}(\hat n)Y_{l_4m_4}(\hat n).
\label{eq:gaunt}
\end{equation}
Given the trispectrum of $\Phi(\vec k)$ of Eq. \eqref{eq:primt}, 
Eq. \eqref{eq:tlm4} leads to
\begin{equation}
t_{l_1l_2l_3l_4}=6g_{\rm NL}\int dr \,r^2\left(
\alpha_{l_1}(r)\beta_{l_2}(r)
\beta_{l_3}(r)\beta_{l_4}(r)
+(\mbox{3 perms}) 
\right), \label{eq:cmbt}
\end{equation}
where $\alpha_l(r)$ and $\beta_l(r)$ are
\begin{eqnarray}
\alpha_l(r)&=&\frac2\pi\int dk\,k^2g_l(k)j_l(kr), 
\label{eq:alpha} \\
\beta_l(r)&=&\frac2\pi\int dk\,k^2g_l(k)j_l(kr)P(k).
\label{eq:beta}
\end{eqnarray}

For later convenience, we introduce a normalized trispectrum
$\hat t_{l_1l_2l_3l_4}$ by
\begin{eqnarray}
\hat t_{l_1l_2l_3l_4}&\equiv& \frac{\partial t_{l_1l_2l_3l_4}}{\partial g_{\rm NL}}\notag \\
&=&6\int dr \,r^2\left(
\alpha_{l_1}(r)\beta_{l_2}(r)
\beta_{l_3}(r)\beta_{l_4}(r)
+(\mbox{3 perms}) 
\right), \label{eq:hatt}
\end{eqnarray}
from which Eq. \eqref{eq:cmbt} can be recast into 
\begin{equation}
t_{l_1l_2l_3l_4}=g_{\rm NL}\hat t_{l_1l_2l_3l_4}.
\label{eq:cmbt2}
\end{equation}

\section{
Real-space representation of $g_{\rm NL}$ estimator
} 
\label{sec:estimator}

According to Refs. \cite{Regan:2010cn,Fergusson:2010gn}, 
an optimal estimator of $g_{\rm NL}$ can be given by
\begin{equation}
\hat g_{\rm NL}= \frac{K_{\rm prim}}{\langle K_{\rm prim}\rangle_{g_{\rm NL}=1}},
\label{eq:estimator}
\end{equation}
where an angle bracket with  subscript $g_{\rm NL}=1$ indicates 
an ensemble average over non-Gaussian simulations with unit $g_{\rm NL}$.
Here $K_{\rm prim}$ is a quartic statistics, which is given by
\begin{eqnarray}
K_{\rm prim}&=&\frac1{24}
\sum_{l_1\cdots l_4}
\sum_{m_1\cdots m_4}
\hat t_{l_1l_2l_3l_4}
\mathcal G^{l_1l_2l_3l_4}_{m_1m_2m_3m_4}
\left[\tilde a_{l_1 m_1}
\tilde a_{l_2 m_2}\tilde a_{l_3 m_3}\tilde a_{l_4 m_4} \right.\notag\\
&&\quad\left.-6\tilde a_{l_1 m_1}\tilde a_{l_2 m_2}
\langle\tilde a_{l_3 m_3}
\tilde a_{l_4 m_4}\rangle_0
+3
\langle\tilde a_{l_1 m_1}\tilde a_{l_2 m_2}\rangle_0
\langle\tilde a_{l_3 m_3}
\tilde a_{l_4 m_4}\rangle_0
\right], 
\label{eq:quartic}
\end{eqnarray}
where the bracket with subscript $0$, $\langle\cdot\rangle_0$, indicates an ensemble average 
over Gaussian simulations. Here
$\tilde a_{lm}$ is a harmonic coefficient obtained from observed (or simulated) data
maps weighted by inverse-variance, 
\begin{equation}
\tilde a_{lm}=\sum_{l'm'}
\mathcal C^{-1}_{lm,l'm'}a_{l'm'},
\label{eq:filtered}
\end{equation}
where $\mathcal C_{lm,l'm'}=C_{lm,l'm'}+N_{lm,l'm'}$ is the total covariance 
with $C_{lm,l'm'}$ and $N_{lm,l'm'}$ being denoted as those of signal and noise, respectively.

As shown in Refs. \cite{Regan:2010cn,Fergusson:2010gn}, 
Eq. \eqref{eq:quartic} is proportional to the
connected part of trispectrum of Eq. \eqref{eq:primt}, so that the 
estimator Eq. \eqref{eq:estimator} is unbiased. 
We can also see that
\begin{eqnarray}
\langle\hat g_{\rm NL}^2\rangle_0
&=&\frac{\langle K_{\rm prim}K_{\rm prim}\rangle_0}{\langle K_{\rm prim}\rangle_{g_{\rm NL}=1}^2} \notag \\
&=&\frac1{\langle K_{\rm prim}\rangle_{g_{\rm NL}=1}}, \label{eq:variance0}
\end{eqnarray}
where we used the relation $
\langle K_{\rm prim}K_{\rm prim}\rangle_0=
\langle K_{\rm prim}\rangle_{g_{\rm NL}=1}$. 
See Appendix \ref{app:norm} for the proof.
Then we finally obtain 
\begin{eqnarray}
\langle \hat g_{\rm NL}^2\rangle_0
&=&\Big[\frac1{24}
\sum_{\substack{l_1\cdots l_4\\l'_1\cdots l'_4}}
\sum_{\substack{m_1\cdots m_4\\m'_1\cdots m'_4}}
\hat t_{l_1l_2l_3l_4}\mathcal G^{l_1l_2l_3l_4}_{m_1m_2m_3m_4}\notag\\
&&\times ~\mathcal C^{-1}_{l_1m_1,l'_1m'_1}\mathcal C^{-1}_{l_2m_2,l'_2m'_2}
\mathcal C^{-1}_{l_3m_3,l'_3m'_3}\mathcal C^{-1}_{l_4m_4,l'_4m'_4}
\hat t_{l'_1l'_2l'_3l'_4}\mathcal G^{l'_1l'_2l'_3l'_4}_{m'_1m'_2m'_3m'_4}\Big]^{-1}. 
\label{eq:variance}
\end{eqnarray}
The right hand side is the inverse of the Fisher matrix
of Eq. \eqref{eq:fisher} with inhomogeneous noise and
sky cuts being taken into account.
Thus Eq. \eqref{eq:variance} shows that Eq. \eqref {eq:estimator} is a
minimum-variance estimator in the limit of weak non-Gaussianity. 

Computation of $K_{\rm prim}$ and $\langle K_{\rm prim}\rangle_{g_{\rm NL}=1}$
should be implemented in the real space rather than the harmonic space.
Combined with Eqs.  \eqref{eq:gaunt} and \eqref{eq:hatt},
Eq. \eqref{eq:quartic} can be rewritten in the real space as
\begin{eqnarray}
K_{\rm prim}&=&
\int d^3r \left[
A(\vec r)B(\vec r)^3
-3A(\vec r)B(\vec r)\langle B(\vec r)^2\rangle_0 \right.\notag \\
&&\left.-3B(\vec r)^2\langle A(\vec r)B(\vec r)\rangle_0
+3\langle A(\vec r)B(\vec r)\rangle_0\langle B(\vec r)^2\rangle_0 
\right], 
\label{eq:real}
\end{eqnarray}
where $A(\vec r)$ and $B(\vec r)$ are defined as
\begin{eqnarray}
A(\vec r)&=&\sum_{lm}\alpha_l(r)Y_{lm}(\hat r)\tilde a_{lm},\\
B(\vec r)&=&\sum_{lm}\beta_l(r)Y_{lm}(\hat r)\tilde a_{lm}.
\end{eqnarray}
The form of Eq. \eqref{eq:real} is computationally
as demanding as the fast estimator of $f_{\rm NL}$ of 
Refs. \cite{Komatsu:2003iq,Yadav:2007rk,Yadav:2007ny}.

\section{Constraint from the WMAP 9-year data}
\label{sec:result}

In order to determine the normalization $\langle K_{\rm prim}\rangle_{g_{\rm NL}=1}$, 
we need to simulate non-Gaussian CMB maps.
We adopt the method for local-type non-Gaussianity 
developed in Ref. \cite{Elsner:2009md}, which uses 
the Gauss quadrature method with optimized nodes and weights 
for line of sight integral. 
Since the method is performed in the real space rather than the harmonic
one, it is straightforwardly extended to the cubic model, 
while it is originally developed for models with up to the quadratic term in Eq. \eqref{eq:NGpert}.
In our analysis, we ask the method to be accurate with mean square of the error less than 0.01 at each multipole $(lm)$.
For $l_{\rm max}=1024$, we found that this level of accuracy requires 42 quadrature nodes.

Computation of inverse-variance weighted map $\tilde a_{lm}$ 
is performed based on the method of Ref. \cite{Smith:2007rg}, 
which uses a conjugate gradient method with multi-grid
preconditioning. This method also allows to marginalize over 
amplitudes of components when their spatial template maps are provided.

The transfer function of CMB is computed using 
the CAMB code \cite{Lewis:1999bs}.
We combine the foreground-cleaned maps of V and W bands of the WMAP 9-year data 
\cite{Bennett:2012fp}.\footnote{http://lambda.gsfc.nasa.gov}
with a resolution $N_{\rm side}=512$ in the HEALPix pixelization scheme 
\cite{Gorski:2004by}.\footnote{http://healpix.jpl.nasa.gov} We adopt the KQ75y9 mask \cite{Bennett:2012fp}
which cuts 31.2\% of the sky. We also set the maximum multipole 
$l_{\rm max}$ to 1024 in our analysis. 
We marginalize the amplitudes of the monopole $l=0$ and dipoles $l=1$ as default
and also optionally marginalize the amplitudes of the Galactic foreground components at large angular
scales using the synchrotron, free-free and dust emission templates from Ref. \cite{Bennett:2012fp}.

Now we are to present our constraints on $g_{\rm NL}$.
Without template marginalization of Galactic foregrounds, 
we obtain $g_{\rm NL}=(-3.9\pm2.2)\times10^5$ at 1 $\sigma$.
With template marginalization, this changes to
$g_{\rm NL}=(-3.3\pm2.2)\times10^5$.
Comparing these two constraints, we see that 
effects of residual Galactic foregrounds are not significant.
Having all these results, we conclude that current CMB data is consistent with Gaussianity.

To validate our analysis, we generated mock WMAP 9-year data from
non-Gaussian CMB simulations with $g_{\rm NL}=10^6$ as well as Gaussian ones with $g_{\rm NL}=0$.
Then we computed the estimator $\hat g_{\rm NL}$ in the same way as the real data but using these mock data.
From the non-Gaussian mock data, we found that our estimator reproduces the input value.
On the other hand, from the Gaussian mock data, we found that the root mean square (rms)
of our estimator found to be consistent with that obtained based on Eq. \eqref{eq:variance0}. 
These results show that our estimation is not biased and self-consistent. 
In addition, as we shall show in the next section, the sizes of the errors are almost the same 
as the expectation from the Fisher matrix analysis, which also supports our result.

Our constraints are a few times stronger than those in 
Refs. \cite{Smidt:2010ra,Fergusson:2010gn}. 
As estimators adopted in these studies are also optimal and in principle the same as ours, 
the improvement should come from differences in the analysis methods. 
While there are a number of differences, we believe that the filtering method and $l_{\rm max}$
affect the constraints most dominantly.
In these references, Eq. \eqref{eq:filtered} is approximated by replacing the full inverse covariance matrix 
$\mathcal C^{-1}_{lm,l'm'}$ with a diagonal one $1/\mathcal C_l\,\delta_{ll'}\delta_{mm'}$.
We therefore repeated the previous validation tests adopting the same mock data but using the approximated 
filtering method. In particular with the Gaussian mock data, we found that the rms of the estimator becomes several times 
larger from one obtained by using the exact filtering method. This may be surprising 
that in the case of $f_{\rm NL}$, use of the suboptimal filtering method increases the error only by a few tens 
of percents (For example, we refer to Refs. \cite{Komatsu:2008hk,Senatore:2009gt} for the case of the WMAP 5-year data). 
Although our test here is far from exhaustive, the result demonstrates that filtering is substantially important in 
optimal estimation of $g_{\rm NL}$.  We also investigated how $l_{\rm max}$ affects the constraints, only to find 
that reducing $l_{\rm max}$ from 1024 to 512 increases the errors only by a few tens of percents.

On the other hand, $g_{\rm NL}$ can be also constrained from the scale-dependent bias in correlation functions
of massive objects. Our constraints are almost comparable to one in Ref. \cite{Giannantonio:2013uqa}, which 
uses data of galaxies and quasars. 
However, there would be one advantage in use of CMB data.
While there is a significant degeneracy between $f_{\rm NL}$ and $g_{\rm NL}$ in constraints from 
the scale-dependent bias as seen in e.g. Ref. \cite{Giannantonio:2013uqa}, there should be no degeneracy between
these two parameters from CMB data at the leading-order both in $f_{\rm NL}$ and $g_{\rm NL}$.
This is because in the limit of Gaussianity, a covariance between a bispectrum and a trispectrum 
of CMB anisotropies should vanish.

\section{Fisher matrix analysis}
\label{sec:fisher}
We here present a method to evaluate an expected 
constraint on $g_{\rm NL}$ based on the Fisher matrix analysis, 
and apply it to the WMAP and forthcoming PLANCK surveys.

Analogously to the case of bispectrum, 
the Fisher matrix for $g_{\rm NL}$ should be approximately given as \cite{Fergusson:2010gn}. 
\begin{equation}
F=\frac{f_{\rm sky}}{24}
\sum_{l_1\cdots l_4}\sum_{m_1\cdots m_4}
\frac{(\hat t_{l_1l_2l_3l_4}
\mathcal G_{m_1m_2m_3m_4}^{l_1l_2l_3l_4})^2}{
\mathcal C_{l_1}\mathcal C_{l_2}
\mathcal C_{l_3}\mathcal C_{l_4}}, 
\label{eq:fisher}
\end{equation}
where $\mathcal C_l=C_l+N_l$ is the total power spectrum and $f_{\rm sky}$ is a sky coverage.
An error of $g_{\rm NL}$ should be given by $1/\sqrt{F}$.

Eq. \eqref{eq:fisher} is computationally expensive as we at least have to carry out
quadruple summation over $l$'s. However, using the fact that the trispectrum $\hat t_{l_1l_2l_3l_4}$ 
of Eq. \eqref{eq:hatt} is in a separable form, Eq. \eqref{eq:fisher} 
can be brought into a computationally cheaper expression as follows:
\begin{eqnarray}
F&=&48\pi^2\int d\mu \int dr~r^2\int dr'~r^{\prime2}
\left[C^{\alpha\alpha}(r,r',\mu)C^{\beta\beta}(r,r',\mu) \right. \\
&&\left.\quad\quad+3C^{\alpha\beta}(r,r',\mu)C^{\beta\alpha}(r,r',\mu)\right] 
{C^{\beta\beta}(r,r',\mu)}^2, \notag
\label{eq:fisher2}
\end{eqnarray}
where $C^{aa'}(r,r',\mu)$, with $a$ and $a'$ being either $\alpha$ or $\beta$, is given by
\begin{equation}
C^{aa'}(r,r',\mu)=\sum_l \frac{(2l+1)a_l(r)a'_l(r')}{4\pi \mathcal C_l}P_l(\mu).
\end{equation}
Here $P_l(\mu)$ is the Legendre polynomial.
Contrary to the quadruple sum over $l$'s of Eq. \eqref{eq:fisher}, 
Eq. \eqref{eq:fisher2} has only a triple integrals, computation of which is thus modest.
Derivation of Eq. \eqref{eq:fisher2} is presented in Appendix \ref{app:fisher}.
We note that a different separable form of the same Fisher matrix 
is presented in Ref. \cite{Fergusson:2010gn}.

We compute $C_l$ using the {\tt CAMB} code \cite{Lewis:1999bs}.
On the other hand, we approximate $N_l$ by the 
Knox's formula \cite{Knox:1995dq}, 
\begin{equation}
N_l=\theta_{\rm FWHM}^2\sigma_T^2\exp\left[l(l+1)\frac{\theta_{\rm FWHM}^2}{8\ln2}\right],
\end{equation}
where $\theta_{\rm FWHM}$ is the 
full width at half maximum of the Gaussian beam, and $\sigma_T$ is the
root mean square of the instrumental noise par pixel.
The total $N_l$ of a multi-frequency survey can be given by 
a quadrature sum of $N_l$ of each frequency band.

First, we estimate the expected error on $g_{\rm NL}$ from the WMAP survey.
Survey parameters we adopted for the WMAP 9-year V and W bands 
for $\theta_{\rm FWHM}$ and $\sigma_T$ are listed in 
Table \ref{tbl:wmap9}.
Furthermore, we assume that $f_{\rm sky}=0.69$, 
which is consistent with the mask we adopt in Section \ref{sec:result}.
From the above setup, we obtain $\Delta g_{\rm NL}=2.1\times10^5$.
The size of error is almost the same as one we obtained from actual WMAP 9-year
data in Section \ref{sec:result}, which supports the validity of our analysis.

\begin{table}[htb]
  \begin{center}
  \begin{tabular}{llcc}
  \hline
  \hline
  band & V & W \\
  \hline
  $\theta_{\rm FWHM}$ [arcmin] & 21.0 & 13.2 \\
  $\sigma_T$ [mK] & 21 & 31 \\
  \hline
  \hline 
\end{tabular}
  \caption{
  Survey parameters for 9-year observation of the WMAP V and W bands \cite{wmap}.
  }
  \label{tbl:wmap9}
\end{center}
\end{table}

Next, we also forecast a constraint from a future survey.
With the survey parameters for PLANCK listed in Table \ref{tbl:planck}
and  $f_{\rm sky}=0.69$, we find that PLANCK will constrain $g_{\rm NL}$ with error 
$\Delta g_{\rm NL}=6.7\times 10^4$, which is a few times tighter than
the current ones. 

\begin{table}[htb]
  \begin{center}
  \begin{tabular}{llccccccc}
  \hline
  \hline
  band &  &  \\
  \hline
  $\theta_{\rm FWHM}$ [arcmin] & 33.0 & 24.0 & 14.0 & 10.0 & 7.1 & 5.0 & 5.0 \\
  $\sigma_T$ [mK] & 5.5 & 7.4 & 12.8 & 6.8 & 6.0 & 13.1 & 40.1 \\
  \hline
  \hline 
\end{tabular}
  \caption{
  Same as in Table \ref{tbl:wmap9} but for the PLANCK 14-month observation
  \cite{Planck:2006aa}.}
  \label{tbl:planck}
\end{center}
\end{table}

\section{Conclusion} \label{sec:conclusion}

We present a method to derive an optimal constraint on 
the non-linearity parameter $g_{\rm NL}$ of
the local-type non-Gaussianity from CMB data.
Computational cost of our method is almost the 
same as the fast $f_{\rm NL}$ estimator of the local-type non-Gaussianity.
Applying our method to the WMAP 9-year data, we obtain 
$g_{\rm NL}=(-3.3\pm2.2)\times 10^5$ with
template marginalization of Galactic foregrounds.
The size of the error is consistent with an expectation based on the Fisher
matrix analysis. 

Our constraints are a few times tighter than ones in
the previous studies \cite{Smidt:2010ra,Fergusson:2010gn} from CMB data.
While the improvement is not very dramatic, 
our constraints are however the first optimal ones on $g_{\rm NL}$ from CMB
with the full covariance matrix being adopted in filtering.
We expect our method can be applied to high resolution data of the PLANCK survey, 
and $g_{\rm NL}$ will be constrained tighter.

In this paper we focused on the cubic term in Eq. \eqref{eq:NGpert}. 
However, higher-order terms in the equation can also be nonzero. We insist that 
our method can be straightforwardly extended for any of the higher-order Gaussian terms 
of the local-type non-Gaussianity as far as we assume that all the other non-Gaussian
terms should vanish.

\bigskip
\bigskip

\noindent 
\section*{Acknowledgment}

This work is supported by Grant-in-Aid for Scientific research from
a Grant-in-Aid for JSPS Research under Grant No. 23-5622 (TS), 
JSPS Grant-in-Aid for Scientific Research under Grant No. 22340056 (NS) 
and Grant-in-Aid for Nagoya University Global COE 
Program ¡ÈQuest for Fundamental Principles in the Universe: from Particles 
to the Solar System and the Cosmos¡É, from the Ministry of Education, 
Culture, Sports, Science and Technology (MEXT) of Japan. 
This research is also supported in part by World Premier International 
Research Center Initiative, MEXT, Japan.
The authors acknowledge Kobayashi-Maskawa Institute for the Origin of
Particles and the Universe, Nagoya University for providing computing
resources  in conducting the research reported in this paper.
Some of the results in this paper have been derived using the HEALPix 
\cite{Gorski:2004by} package.

\appendix

\noindent 
\section{Equivalence of $\langle K_{\rm prim}K_{\rm prim}\rangle_0$ and
$\langle K_{\rm prim}\rangle_{g_{\rm NL}=1}$ }
\label{app:norm}
In this appendix, we present a proof for $\langle K_{\rm prim}K_{\rm prim}\rangle_0
=\langle K_{\rm prim}\rangle_{g_{\rm NL}=1}$. 
First, we divide the CMB anisotropy $a_{lm}$ into the Gaussian $a_{{\rm G},lm}$ and
non-Gaussian $a_{{\rm NG},lm}$ parts:
\begin{eqnarray}
a_{{\rm G},lm} &=&4\pi(-i)^l\int \frac{d^3k}{(2\pi)^3}g_l(k)\Phi_{\rm G}(\vec k)Y^*_{lm}(\hat k), \\
a_{{\rm NG},lm} &=&4\pi(-i)^l\int \frac{d^3k}{(2\pi)^3}g_l(k)\Phi_{\rm G}^3(\vec k)Y^*_{lm}(\hat k), 
\end{eqnarray}
which leads to $a_{lm}=a_{{\rm G},lm}+g_{\rm NL}a_{{\rm NG},lm}$.
Since filtered map $\tilde a_{lm}$ is a linear function of $a_{lm}$, 
$\tilde a_{lm}$ can also be divided into the Gaussian and non-Gaussian parts in the same way, 
$\tilde a_{lm}=\tilde a_{{\rm G},lm}+g_{\rm NL}\tilde a_{{\rm NG},lm}$.

First we compute $\langle K_{\rm prim}\rangle_{g_{\rm NL}=1}$.
Up to the leading order in $a_{{\rm NG},lm}$, 
Eq. \eqref{eq:quartic} leads to
\begin{eqnarray}
\langle K_{\rm prim}\rangle_{g_{\rm NL}=1}
&=&\frac1{24}
\sum_{l_1\cdots l_4}\sum_{m_1\cdots m_4}
\hat t_{l_1l_2l_3l_4}\mathcal G^{l_1l_2l_3l_4}_{m_1m_2m_3m_4}
\left[4\langle\tilde a_{{\rm NG},l_1m_1}\tilde a_{{\rm G},l_2m_2}
\tilde a_{{\rm G},l_3m_3}\tilde a_{{\rm G},l_4m_4}\rangle \right.\notag\\
&&\quad\quad\left.-12\langle \tilde a_{{\rm NG},l_1m_1}\tilde a_{{\rm G},l_2m_2}\rangle
\langle \tilde a_{{\rm G},l_3m_3}\tilde a_{{\rm G},l_4m_4}\rangle\right] \notag\\
&=&\frac1{24}
\sum_{\substack{l_1\cdots l_4\\l'_1\cdots l'_4}}
\sum_{\substack{m_1\cdots m_4\\m'_1\cdots m'_4}}
\hat t_{l_1l_2l_3l_4}\mathcal G^{l_1l_2l_3l_4}_{m_1m_2m_3m_4}\label{eq:K_g1}\\
&&\times ~\mathcal C^{-1}_{l_1m_1,l'_1m'_1}\mathcal C^{-1}_{l_2m_2,l'_2m'_2}
\mathcal C^{-1}_{l_3m_3,l'_3m'_3}\mathcal C^{-1}_{l_4m_4,l'_4m'_4}
\hat t_{l'_1l'_2l'_3l'_4}\mathcal G^{l'_1l'_2l'_3l'_4}_{m'_1m'_2m'_3m'_4}, \notag
\end{eqnarray}
where we used the relation
\begin{eqnarray}
\langle a_{l_1m_1} a_{l_2m_2}a_{l_3m_3}a_{l_4m_4}
\rangle_{\rm conn}&=&g_{\rm NL}\left[\{
\langle a_{{\rm NG},l_1m_1}\tilde a_{{\rm G},l_2m_2}
\tilde a_{{\rm G},l_3m_3}\tilde a_{{\rm G},l_4m_4}\rangle
+\mbox{(3 perms)}\} \right.\\
&&\left.\quad\quad-\{\langle \tilde a_{{\rm NG},l_1m_1}\tilde a_{{\rm G},l_2m_2}\rangle
\langle \tilde a_{{\rm G},l_3m_3}\tilde a_{{\rm G},l_4m_4}\rangle
+\mbox{(11 perms)}\}
\right],\notag
\end{eqnarray}
and
\begin{equation}
\langle \tilde a_{{\rm G},lm}\tilde a^*_{{\rm G},l'm'} \rangle
=\mathcal C^{-1}_{lm,l'm'}. \label{eq:invcov}
\end{equation}

On the other hand, 
$\langle K_{\rm prim}K_{\rm prim}\rangle_0$ is also computed from
Eq. \eqref{eq:quartic}, which leads to
\begin{eqnarray}
\langle K_{\rm prim}K_{\rm prim}\rangle_0&=&
\frac{1}{24^2}
\sum_{\substack{l_1\cdots l_4\\l'_1\cdots l'_4}}
\sum_{\substack{m_1\cdots m_4\\m'_1\cdots m'_4}}
\hat t_{l_1l_2l_3l_4}G^{l_1l_2l_3l_4}_{m_1m_2m_3m_4}
\hat t_{l'_1l'_2l'_3l'_4}G^{l'_1l'_2l'_3l'_4}_{m'_1m'_2m'_3m'_4} \\
&&\times\langle \Big[\tilde a_{{\rm G},l_1m_1}\tilde a_{{\rm G},l_2m_2}
\tilde a_{{\rm G},l_3m_3}\tilde a_{{\rm G},l_4m_4}
-6\tilde a_{{\rm G},l_1m_1}\tilde a_{{\rm G},l_2m_2}
\langle\tilde a_{{\rm G},l_3m_3}\tilde a_{{\rm G},l_4m_4}\rangle \notag\\
&&\quad+3\langle\tilde a_{{\rm G},l_1m_1}\tilde a_{{\rm G},l_2m_2}\rangle
\langle\tilde a_{{\rm G},l_3m_3}\tilde a_{{\rm G},l_4m_4}\rangle
\Big]
\Big[\tilde a^*_{{\rm G},l'_1m'_1}\tilde a^*_{{\rm G},l'_2m'_2}
\tilde a^*_{{\rm G},l'_3m'_3}\tilde a^*_{{\rm G},l'_4m'_4}\notag\\
&&\quad-6\tilde a^*_{{\rm G},l'_1m'_1}\tilde a^*_{{\rm G},l'_2m'_2}
\langle\tilde a^*_{{\rm G},l'_3m'_3}\tilde a^*_{{\rm G},l'_4m'_4}\rangle
+3\langle\tilde a^*_{{\rm G},l'_1m'_1}\tilde a^*_{{\rm G},l'_2m'_2}\rangle
\langle\tilde a^*_{{\rm G},l'_3m'_3}\tilde a^*_{{\rm G},l'_4m'_4}\rangle
\Big]\rangle. \notag 
\end{eqnarray}
By using the Wick theorem, after lengthy but simple calculation, 
we obtain
\begin{eqnarray}
\langle K_{\rm prim}K_{\rm prim}\rangle_0&=&
\frac1{24}
\sum_{\substack{l_1\cdots l_4\\l'_1\cdots l'_4}}
\sum_{\substack{m_1\cdots m_4\\m'_1\cdots m'_4}}
\hat t_{l_1l_2l_3l_4}\mathcal G^{l_1l_2l_3l_4}_{m_1m_2m_3m_4}
\label{eq:KK0}\\
&&\times ~\mathcal C^{-1}_{l_1m_1,l'_1m'_1}\mathcal C^{-1}_{l_2m_2,l'_2m'_2}
\mathcal C^{-1}_{l_3m_3,l'_3m'_3}\mathcal C^{-1}_{l_4m_4,l'_4m'_4}
\hat t_{l'_1l'_2l'_3l'_4}\mathcal G^{l'_1l'_2l'_3l'_4}_{m'_1m'_2m'_3m'_4}.
\notag
\end{eqnarray}

Comparison of Eqs. \eqref{eq:K_g1} and \eqref{eq:KK0} shows 
$\langle K_{\rm prim}K_{\rm prim}\rangle_0=\langle K_{\rm prim}\rangle_{g_{\rm NL}=1}$.

\noindent 
\section{Derivation of Fisher matrix}
\label{app:fisher}
In this appendix, we derive Eq. \eqref{eq:fisher2}.
We first start from computation of the summation over $m$'s in Eq. \eqref{eq:fisher}.
Using Eq. \eqref{eq:gaunt}, we obtain
\begin{eqnarray}
\sum_{\{m\}}
(\mathcal G^{l_1l_2l_3l_4}_{m_1m_2m_3m_4})^2
&=&\int d\hat n\int d\hat n' 
\prod^{4}_{i=1}[\sum_{m_i} Y_{l_im_i}(\hat n) Y_{l_im_i}^*(\hat n')] \\
&=&\frac{(2l_1+1)(2l_2+1)(2l_3+1)(2l_4+1)}{32\pi^2}\int d\mu
P_{l_1}(\mu)P_{l_2}(\mu)P_{l_3}(\mu)P_{l_4}(\mu), \notag
\label{eq:g2}
\end{eqnarray}
where in the second equality we used the relation
\begin{equation}
\sum_m Y_{lm}(\hat n) Y_{lm}^*(\hat n')=
\frac{2l+1}{4\pi}P_l(\hat n\cdot \hat n').
\end{equation}
Then adopting the definition of $\hat t_{l_1l_2l_3l_4}$
of Eq. \eqref{eq:hatt}, Eq. \eqref{eq:fisher} can be rewritten
as
\begin{eqnarray}
F=12\pi^2\int dr\,r^2 \int dr'\,r^{\prime2}\int d\mu
\sum_{\vec a,\vec a'} \prod_{i=1}^4\left[\frac{(2l+1)a_l^{(i)}(r)a_l^{\prime\,(i)}(r')}
{4\pi \mathcal C_l}P_l(\mu)\right],
\end{eqnarray}
where $a_l^{(i)}(r)$ and $a_l^{\prime\,(i)}(r)$ should be 
either $\alpha_l(r)$ and $\beta_l(r)$.
We also introduced a vector $\vec a=(a^{(1)},\,a^{(2)},\,a^{(3)},\,a^{(4)})$, 
which symbolically represents a permutation of $(\alpha, \beta, \beta, \beta)$.
By taking summations over $\vec a$ and $\vec a^\prime$ of any possible
permutations, then we finally obtain Eq. \eqref{eq:fisher2}.


\end{document}